\documentstyle[twoside,fleqn,espcrc2,epsfig]{article}

\newcommand{\AmS}{{\protect\the\textfont2
  A\kern-.1667em\lower.5ex\hbox{M}\kern-.125emS}}

\newcommand{\be}{\begin{equation}}
\newcommand{\ee}{\end{equation}}
\newcommand{\bea}{\begin{eqnarray}}
\newcommand{\eea}{\end{eqnarray}}

\newcommand{\ep}{\varepsilon}
\newcommand{\nn}{\nonumber}

\newcommand{\pslash}{p\hspace{-2mm}/}
\newcommand{\half}{{\textstyle\frac{1}{2}}}
\newcommand{\quarter}{{\textstyle\frac{1}{4}}}

\title{
$\qquad\qquad\qquad\qquad\qquad\qquad\qquad\qquad\qquad
 \qquad\qquad\qquad\quad$ MZ-TH/00-23\\[2mm]
One-loop results for the quark-gluon vertex 
in arbitrary dimension\thanks{Based on the talk
given by P.~O. at the Zeuthen Workshop ``Loops and Legs in Gauge Theories''
(Bastei, Germany, April 2000).
Published in Nucl. Phys. B (Proc. Suppl.) 89 (2000) 277-282.}}

\author{A.~I.~Davydychev\address{Department of Physics,
University of Mainz,
Staudingerweg 7,
D-55099 Mainz, Germany}%
\thanks{On leave from 
 Institute for Nuclear Physics, Moscow State University, 
 119899, Moscow, Russia.},
P.~Osland${}^{\rm b}$,
and
L.~Saks\address{Department of Physics, University of Bergen,
        All\'{e}gaten 55, N-5007 Bergen, Norway},
}

\begin{document}

\begin{abstract}
Results on the one-loop quark-gluon vertex with massive quarks are
reviewed, in an arbitrary covariant gauge and in arbitrary space-time
dimension. We show how it is possible to get on-shell results from the
general off-shell expressions. The corresponding Ward-Slavnov-Taylor
identity is discussed.
\end{abstract}

\maketitle

\section{INTRODUCTION}

The quark-gluon vertex is one of the fundamental objects in Quantum
Chromodynamics (QCD) \cite{QCD}. To calculate various higher-order QCD
effects we should have a detailed knowledge of this
vertex.  For the one-loop quark-gluon vertex we have two
contributions.  
The ``Abelian'' contribution is equal, up to an overall factor,
to the fermion-photon vertex in QED.
In the four-dimensional case,
off-shell results for the latter have been obtained in \cite{BC1}
(Feynman gauge) and \cite{KRP} (an arbitrary covariant gauge).
In three dimensions, results for the massless case are 
available in \cite{BKP}. 

In special limits some results are also available for the second, 
non-Abelian contribution. 
In Ref.~\cite{PT} (see also \cite{CG}), the ``symmetric'' case, 
$p_1^2=p_2^2=p_3^2$, was treated in an arbitrary covariant gauge, 
for massless quarks. 
For massive quarks, this $p_1^2=p_2^2=p_3^2$ case was examined
in \cite{DTP} (only the coefficient of $\gamma_{\mu}$ was 
calculated).
Some configurations when the gluon (or quark)
momentum vanishes
were considered in~\cite{BL}. Some
on-shell results are available for massless quarks \cite{NPS}. 
Recently we completed a study of the one-loop quark-gluon vertex in
an arbitrary covariant gauge and space-time dimension \cite{DOS}. 

\section{FORMALISM}

The lowest-order quark-gluon vertex is
\begin{equation}
\label{qg_vert} 
g(T^{a})_{ji}\left[\gamma_{\mu}\right]_{\beta \alpha} ,
\end{equation}
where $T^{a}$ are colour matrices corresponding to the
fundamental representation of the gauge group.
We will consider the SU($N$) group, with $N$ the number of colours.
Extracting the over-all colour structure, we can present the 
``dressed'' (one-particle irreducible) quark-gluon vertex as
\begin{equation}
\label{qqg}
\Gamma_{\mu}^{a}(p_1,p_2,p_3)
= g T^{a}\, \Gamma_{\mu}(p_1,p_2,p_3) ,
\end{equation}
where $p_1$, $p_2$ are momenta of the quarks, $p_3$ is the gluon momentum, 
all of which are ingoing, $p_1+p_2+p_3=0$.

At the one-loop level, we have two contributions to the
quark-gluon vertex (see Fig.~1). Their colour factors are 
proportional to $\left(C_F-\half C_A\right)$ and
$C_A$, where $C_F$ and $C_A$ denote
eigenvalues of the quadratic Casimir operator in
the fundamental and adjoint representations, respectively.
For the SU($N$) gauge group,
$C_A=N$, and
$C_F=(N^2-1)/(2N)$. 
\begin{figure}[t]
\refstepcounter{figure}
\label{Fig:1}
\addtocounter{figure}{-1}
\begin{center}
\setlength{\unitlength}{1cm}
\begin{picture}(4,2.5)
\put(-4.5,-2.0){
\mbox{\epsfysize=5.5cm\epsffile{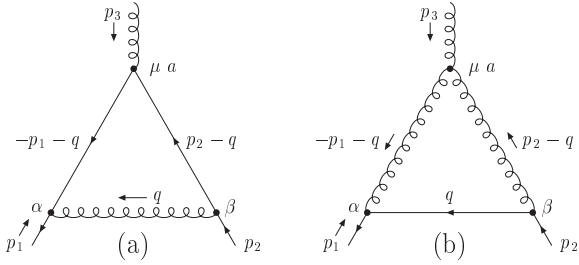}}}
\end{picture}
\caption{
The two one-loop diagrams.
}
\end{center}
\vspace*{-8mm}
\end{figure}

\subsection{Ward-Slavnov-Taylor (WST) identity}

The WST identity \cite{WST} for the quark-gluon vertex 
$\Gamma_{\mu}(p_1,p_2,p_3)$
reads (see, e.g., in \cite{PT-QCD})
\begin{eqnarray}
\label{WST_qqg}
p_3^{\mu} \Gamma_{\mu}(p_1,p_2,p_3)
\hspace*{-3mm} &=& \hspace*{-3mm} G(p_3^2)\left[ S^{-1}(-p_1)\; H(p_1,p_2,p_3)\right.
\nn \\
&&-\left. \overline{H}(p_2,p_1,p_3)\; S^{-1}(p_2) \right],
\end{eqnarray}
where $G(p^2)$ is the scalar function associated 
with the ghost propagator (the latter is proportional
to $G(p^2)/p^2$);
$S^{-1}$ is the inverse quark propagator; 
the function $H$ (and the ``conjugated'' function $\overline{H}$)
involves the complete four-point quark-quark-ghost-ghost vertex
(for details, see \cite{PT-QCD,DOS}).

At the one-loop level, it is convenient to ``split'' the 
WST identity into two separate relations, corresponding
to the contributions of the two diagrams.
To do this, we need to rewrite the one-loop contribution
to the r.h.s.\ of (\ref{WST_qqg}) in terms of colour
coefficients $(C_F-\half C_A)$ and $C_A$, by analogy with
the two contributions to the l.h.s. On the r.h.s., all 
one-loop contributions are proportional to $C_A$, 
except for the quark self energies, which
contain $C_F$. Therefore, all we need to do is to
represent this $C_F$ as $(C_F-\half C_A)+\half C_A$.
In this way, we get two separate WST identities for
the contributions of diagrams $a$ and $b$ in Fig.~1\footnote{
For a quantity $X$, we denote the zero-loop-order
contribution as $X^{(0)}$, and the one-loop-order contribution as
$X^{(1)}$, so that the perturbative expansion looks like
$X = X^{(0)} + X^{(1)} + \ldots$.},
\begin{eqnarray}
\label{WST1}
&&\hspace*{-7mm}p_3^{\mu} \Gamma_{\mu}^{(1a)}(p_1,p_2,p_3)
= 
\left(C_F-\half C_A\right)\; C_F^{-1} 
\nn \\ 
&\hspace*{-6mm}& \hspace{-5mm} \times
\left[ S^{-1}(-p_1) - S^{-1}(p_2) \right]^{(1)} ,
\\
\label{WST2}
&&\hspace*{-7mm}p_3^{\mu} \Gamma_{\mu}^{(1b)}(p_1,p_2,p_3)
= \left[ S^{-1}(-p_1) \right]^{(0)} \! H^{(1)}(p_1,p_2,p_3)
\nn \\
&\hspace*{-6mm} - & \hspace{-5mm}{\overline{H}}^{(1)}(p_2,p_1,p_3)\; 
     \left[ S^{-1}(p_2) \right]^{(0)}
\nn \\
 &\hspace*{-6mm} + & \hspace{-5mm} \half C_A \; C_F^{-1} \;
\left[ S^{-1}(-p_1) - S^{-1}(p_2) \right]^{(1)}\; H^{(0)}
\nn \\
 &\hspace*{-6mm}+ & \hspace{-5mm} 2 G^{(1)}(p_3^2)\; 
\left[ S^{-1}(-p_1) - S^{-1}(p_2) \right]^{(0)}\; H^{(0)} .
\end{eqnarray}
The first identity has (up to a factor) the same structure
as the Abelian (QED) identity. The second identity
is a ``non-Abelian'' one.

\subsection{Decomposition of the vertex}

Because of the WST identity (\ref{WST_qqg}), it is convenient to decompose
the quark-gluon vertex into a longitudinal part and a transverse part,
as proposed in \cite{BC1},
\begin{equation}
\label{LT}
\Gamma_{\mu}\!(p_1,p_2,p_3)\!
= \!\Gamma_{\mu}^{({\rm L})}\!(p_1,p_2,p_3) 
\!+\! \Gamma_{\mu}^{({\rm T})}\!(p_1,p_2,p_3)
\end{equation}
where
\begin{equation}
\label{GammaL}
\Gamma_{\mu}^{({\rm L})}(p_1,p_2,p_3)
\! = \! \sum\limits_{i=1}^4 \! \lambda_i(p_1^2,p_2^2,p_3^2)\;
\! L_{i,\mu}(p_1,p_2) ,
\end{equation}
\begin{equation}
\label{GammaT}
\Gamma_{\mu}^{({\rm T})}(p_1,p_2,p_3)
\! = \!\sum\limits_{i=1}^8\!
\tau_{i}(p_1^2,p_2^2,p_3^2) T_{i,\mu}(p_1,p_2),
\end{equation}
with
\begin{eqnarray}
\label{L_i}
L_{1,\mu} &=& \gamma_{\mu},
\nn \\
L_{2,\mu} &=& (\pslash_1-\pslash_2)(p_1-p_2)_{\mu},
\nn \\
L_{3,\mu} &=& (p_1-p_2)_{\mu},
\nn \\
L_{4,\mu} &=& \sigma_{\mu \nu}\, (p_1-p_2)^{\nu} ,
\\
\label{T_i}
T_{1,\mu}
&=& p_{1\mu}(p_2 p_3)-p_{2\mu}(p_1 p_3),
\nn \\
T_{2,\mu}&=&
\left[-p_{1\mu}(p_2 p_3)+p_{2\mu}(p_1 p_3)\right](\pslash_1-\pslash_2),
\nn \\
T_{3,\mu}&=&p_3^2\gamma_{\mu}-p_{3\mu}\pslash_3,
\nn \\
T_{4,\mu}&=&\left[p_{1\mu}(p_2 p_3)-p_{2\mu}(p_1 p_3)\right]
\sigma_{\nu \lambda}\, p_1^{\nu}p_2^{\lambda},
\nn \\
T_{5,\mu}&=& \sigma_{\mu \nu}\, p_3^\nu,
\nn \\
T_{6,\mu}&=&\gamma_{\mu} (p_1^2-p_2^2)+(p_1-p_2)_{\mu}\pslash_3,
\nn \\
T_{7,\mu}&=&\half
(p_2^2-p_1^2)\left[\gamma_{\mu}(\pslash_1-\pslash_2)-(p_1-p_2)_{\mu}\right]
\nn \\
&&-(p_1-p_2)_{\mu}\sigma_{\nu\lambda}\, p_1^{\nu}p_2^{\lambda},
\nn \\
T_{8,\mu}&=&
-\gamma_{\mu}\sigma_{\nu\lambda}\, p_1^{\nu}p_2^{\lambda}
+p_{1\mu}\pslash_2-p_{2\mu}\pslash_1.
\end{eqnarray}
The transverse part $\Gamma_{\mu}^{({\rm T})}$ does not contribute to the
WST identity (\ref{WST_qqg}),
$p_3^{\mu}\Gamma_{\mu}^{({\rm T})}(p_1,p_2,p_3)\!=\!0$.

Furthermore, in Ref.~\cite{KRP} a modification of the basis 
(\ref{GammaT})--(\ref{T_i}) has been proposed, which has an advantage
in dealing with kinematical singularities\footnote{In Ref.~\cite{KRP}, the
notation $\sigma_i$ and $S_i$ was used for what we call 
$\widetilde{\tau}_i$ and $\widetilde{T}_i$.}. 
Namely, the transverse part is represented as
\begin{eqnarray}
\label{GammaTtilde}
\Gamma_{\mu}^{({\rm T})}\!(p_1,p_2,p_3)\!=\!\sum\limits_{i=1}^8
\!\widetilde{\tau}_i(p_1^2,p_2^2,p_3^2) \widetilde{T}_{i,\mu}(p_1,p_2),
\end{eqnarray}
where
\begin{eqnarray}
\widetilde{T}_{4,\mu}
= \frac{2}{p_2^2-p_1^2}\left[ 2T_{4,\mu} - p_3^2 T_{7,\mu} \right],
\\
\widetilde{\tau}_4 = \quarter(p_2^2-p_1^2) \tau_4, \qquad
\label{tau7t}
\widetilde{\tau}_7=\tau_7+\half p_3^2 \tau_4 ,
\end{eqnarray}
whereas $\widetilde{T}_{i,\mu}=T_{i,\mu} \;\; (i\neq 4)$ and
$\widetilde{\tau}_i=\tau_i \;\; (i=1,2,3,5,6,8)$.
Moreover, in the on-shell limit the following
modifications of $\lambda_2$ and $\lambda_3$ turn out to be useful:
\begin{equation}
\label{lambda23t}
\widetilde{\lambda}_2 \equiv \lambda_2 +\half p_3^2\tau_2,
\hspace{10mm}
\widetilde{\lambda}_3 \equiv \lambda_3 -\half p_3^2\tau_1 \; .
\end{equation}

Applying charge conjugation to the quark-gluon vertex, 
i.e., interchanging quark and anti-quark,
one obtains 
(see, e.g., in Ref.~\cite{KRP}):
\begin{eqnarray}
\label{c_vertex_c_inv}
{\rm C}\;\Gamma_{\mu}(p_1,p_2,p_3)\;{\rm C}^{-1}
=-\Gamma_{\mu}^{\rm T}(p_2,p_1,p_3).
\end{eqnarray}
Interchanging the quark momenta ($p_1 \leftrightarrow p_2$)
and using the fact that
${\rm C}\;\gamma_{\mu}\;{\rm C}^{-1}=-\gamma_{\mu}^{\rm T}$,
one finds that all $L_\mu$ and $T_\mu$ 
are odd, except for $L_{4,\mu}$ and $T_{6,\mu}$, which are even.
Thus, to satisfy Eq.~(\ref{c_vertex_c_inv}) all $\lambda$'s and $\tau$'s
must be symmetric under the interchange of $p_1^2$ and $p_2^2$, except 
$\lambda_4$ and $\tau_6$, which are odd.
An important corollary is that in the
case $p_1^2=p_2^2\equiv p^2$ the $\lambda_4$ and $\tau_6$
functions must vanish,
\begin{equation}
\label{l4t6}
\lambda_4(p^2,p^2,p_3^2) = 0, \qquad
\tau_6(p^2,p^2,p_3^2) = 0 \; .
\end{equation}

\subsection{Integrals}

The two one-loop quark-gluon vertex diagrams involve two triangle
integrals:
\begin{eqnarray}
\label{defJ2}
&& \hspace*{-7mm}
J_2(\nu_1,\nu_2,\nu_3)
\nn \\
&& \hspace*{-7mm}
\equiv 
\int \! \!
 \frac{\mbox{d}^n q}{ \left[(p_2 \!-\!q )^2\!-\!m^2\right]^{\nu_1} \!
                      \left[(p_1 \!+\!q )^2\!-\!m^2\right]^{\nu_2} \! 
      (q^2)^{\nu_3} } ,
\end{eqnarray}
\begin{eqnarray}
\label{defJ1}
&& \hspace*{-7mm}
J_1(\nu_1,\nu_2,\nu_3)
\nn \\
&& \hspace*{-7mm}
\equiv 
\int
 \frac{\mbox{d}^n q}{ \left[(p_2 -q )^2\right]^{\nu_1}  
                      \left[(p_1 +q )^2\right]^{\nu_2}
      \left[q^2-m^2\right]^{\nu_3} } ,
\end{eqnarray}
where $n=4-2\ep$ is the space-time dimension in the   
framework of dimensional regularization \cite{dimreg}.
The subscript of $J$ indicates the number of massive
propagators (see, e.g., in \cite{BD-TMF}).

Tensor integrals can be reduced to the scalar
ones using the standard techniques \cite{BF+PV}
(see also in \cite{PLB'91,Tar}).
Using the integration-by-parts technique \cite{ibp},
all integrals with higher integer powers 
of propagators can be algebraically reduced to integrals
with the powers equal to one or zero
(for details, see \cite{JPA}).

Two non-trivial integrals appear\footnote{
We have extended the notation used in \cite{DOT1} 
by introducing the functions $\varphi_i$ ($i=1,2$),
where $i$ shows the number of massive propagators involved.},
\begin{equation}
\label{J_i111}
J_i(1,1,1)={\mbox{i}}\; \pi^{n/2}\; \eta\; \varphi_i(p_1^2,p_2^2,p_3^2;m),
\end{equation}
where we have extracted a factor
\begin{equation}
\label{eta}
\eta \equiv \!
\frac{\Gamma^2(\frac{n}{2}-1)}{\Gamma(n-3)} 
     \Gamma(3-{\textstyle{n\over2}}) =\!
\frac{\Gamma^2(1-\varepsilon)}{\Gamma(1-2\varepsilon)} 
\Gamma(1+\varepsilon) .
\end{equation}
Then, 
the following two-point functions appear:
\begin{eqnarray}
J_1(1,1,0)=J_0(1,1,0)={\mbox{i}}\; \pi^{n/2}\; \eta\; \kappa_{0,3} \; ,
\nn \\
J_1(0,1,1)=J_2(0,1,1)={\mbox{i}}\; \pi^{n/2}\; \eta\; \kappa_{1,1} \; , 
\nn \\
J_1(1,0,1)=J_2(1,0,1)={\mbox{i}}\; \pi^{n/2}\; \eta\; \kappa_{1,2} \; .
\end{eqnarray}
We have introduced the functions 
$\kappa_i(p_l^2;m)\equiv \kappa_{i,l}$, 
where $p_l$ ($l=1,2,3$) is the external momentum of the two-point
function, whereas the subscript $i=0,1,2$ shows how many
of the two internal propagators are massive.
For massive lines we can also get ``tadpole'' integrals:
\begin{eqnarray}
&&\hspace*{-7mm} J_1(0,0,1)=J_2(1,0,0)=J_2(0,1,0)\nn \\
&&\hspace*{-5mm}= 
{\mbox{i}}\; \pi^{n/2}
\frac{\Gamma(1+\ep)}{\ep(1-\ep)} (m^2)^{1-\ep}
= {\mbox{i}}\; \pi^{n/2} \;\eta\; m^2 \; \widetilde{\kappa} ,
\hspace{4mm}
\end{eqnarray}
with 
\begin{equation}
\widetilde{\kappa} \equiv\widetilde{\kappa}(m^2)
\equiv \frac{\Gamma(1-2\ep)}{\Gamma^2(1-\ep)}\;
\frac{1}{\ep(1-\ep)}\; (m^2)^{-\ep} .
\end{equation}
Massless tadpoles vanish in the framework of dimensional
regularization \cite{dimreg}.
In four dimensions, the integrals (\ref{J_i111}) can be evaluated
in terms of dilogarithms \cite{tHV'79}.

\section{OFF-SHELL RESULTS}

Before presenting selected results for the $\lambda$ and $\tau$
functions, let us 
introduce the following notation for the occurring Gram determinants:
\begin{eqnarray}
\label{Eq:calK}
{\cal{K}} \! &\equiv& \! p_1^2 p_2^2 - (p_1 p_2)^2 \; ,
\\
{\cal{M}}_1 \! &\equiv& \! 
{\cal{K}} + \left( (p_1 p_2)+m^2 \right)^2 \; ,
\\
{\cal{M}}_2 \! &\equiv& \!
(p_1^2\!-\!m^2)(p_2^2\!-\!m^2)p_3^2\!+\!m^2 (p_1^2\!-\!p_2^2)^2 \; .
\end{eqnarray}

We consider the contributions of diagrams $a$ and $b$
(see Fig.~1) separately,
\begin{equation}
\lambda_i^{(1)} = \lambda_i^{(1a)} + \lambda_i^{(1b)} \; ,
\hspace{3mm}
\tau_i^{(1)} = \tau_i^{(1a)} + \tau_i^{(1b)} \; .
\end{equation}

General results for the longitudinal parts of the vertex are
reasonably compact, even in a general covariant gauge and arbitrary
dimension (see in \cite{DOS}).
Below, we show a few selected results.

\subsection{Longitudinal part}

The $\lambda$ functions for diagram $a$ are 
\begin{eqnarray}
\lambda_1^{(1a)}(p_1^2,p_2^2,p_3^2) \hspace*{-3mm} &=& \hspace*{-3mm} 
\frac{g^2\eta \left(C_F\!-\!\half C_A\right)}{(4\pi)^{n/2}} 
      \frac{(n\!-\!2)(1\!-\!\xi)}{4 p_1^2 p_2^2}
\nonumber  \\ 
&\hspace*{-35mm}& \hspace{-30mm} \times \!\!
\left[ p_1^2 (p_2^2\!\!+\!\!m^2) \kappa_{1,2}\!+\!p_2^2 (p_1^2\!\!+\!\!m^2) 
\kappa_{1,1}
         \!-\!(p_1^2\!\!+\!\!p_2^2) m^2 \! \widetilde{\kappa} \right],
\nonumber  \\
\lambda_2^{(1a)}(p_1^2,p_2^2,p_3^2) \hspace*{-3mm} &=& \hspace*{-3mm}
\frac{g^2\eta \left(C_F\!-\!\half C_A\right)}{(4\pi)^{n/2}}
\frac{(n\!-\!2)(\xi\!-\!1)}{4 p_1^2 p_2^2 (p_1^2\!-\!p_2^2)}
\nonumber  \\
&\hspace*{-35mm}& \hspace{-30mm} \times \!\!
\left[ p_1^2 (p_2^2\!\!+\!\!m^2) \kappa_{1,2}\! 
      -\! p_2^2(p_1^2\!\!+\!\!m^2) \kappa_{1,1}
         \!-\!(p_1^2\!\!-\!\!p_2^2) m^2\! \widetilde{\kappa} \right],
\nonumber  \\
\lambda_3^{(1a)}(p_1^2,p_2^2,p_3^2) \hspace*{-3mm} &=& \hspace*{-3mm} 
- \frac{g^2\eta \left(C_F\! -\! \half C_A\right)}{(4\pi)^{n/2}}
\frac{(n\! -\! \xi) m}{p_1^2\!-\!p_2^2} \nonumber  \\
&\hspace*{-8mm}& \hspace{-3mm} \times \!\!
\left[ \kappa_{1,2} - \kappa_{1,1} \right] ,
\nonumber  \\
\lambda_4^{(1a)}(p_1^2,p_2^2,p_3^2) \hspace*{-3mm} &=& \hspace*{-3mm}  0 .
\end{eqnarray}

For diagram $b$, the results are a bit longer. Here we present
the $\lambda$'s in Feynman gauge:
\begin{eqnarray}
\left.\lambda_1^{(1b)}\!(p_1^2,p_2^2,p_3^2)\!\right|_{\xi=0} 
\hspace*{-5mm} &=& \hspace*{-3mm}
- \frac{g^2\eta C_A}{(4\pi)^{n/2}} \! \frac{1}{8} \! 
\biggl[ 
 2  (p_1^2 \!\!+\!\! p_2^2 \!\!- \!\!2  m^2)  \varphi_1  
\nonumber \\ 
&&-
4 \kappa_{0,3}-n  \kappa_{1,2} -n  \kappa_{1,1} 
\nonumber \\ 
&\hspace*{-55mm} + & \hspace{-30mm}
\frac{(p_1^2-p_2^2)}{{\cal{K}}}  
\Bigl( 
(p_1^2-p_2^2) 
\{ 
[(p_1 p_2)+m^2]  \varphi_1 + \kappa_{0,3}  \}
\nonumber \\ 
&\hspace*{-25mm} + & \hspace{-15mm}
[p_2^2-(p_1 p_2)]  \kappa_{1,2} -[p_1^2-(p_1 p_2)]  \kappa_{1,1}  \Bigr)
\nonumber \\
&\hspace*{-55mm} - & \hspace{-30mm}
(n-2) \biggl(\frac{m^2}{p_2^2} \kappa_{1,2}+\frac{m^2}{p_1^2}
\kappa_{1,1}
-\frac{p_1^2+p_2^2}{p_1^2 p_2^2} m^2 \widetilde{\kappa}  \biggr)
 \biggr],
\nonumber \\
\left.\lambda_2^{(1b)}\!(p_1^2,p_2^2,p_3^2)\!\right|_{\xi=0} 
\hspace*{-5mm} &=& \hspace*{-3mm}
\frac{g^2\eta C_A}{(4\pi)^{n/2}}\! \frac{1}{8}\!
  \biggl[  
\frac{p_3^2}{{\cal{K}}} 
\biggl( 
[(p_1 p_2)\!+\!m^2] \varphi_1 
\nonumber \\ 
&\hspace*{-55mm} + & \hspace{-30mm} 
 \kappa_{0,3}  
+\frac{p_2^2-(p_1 p_2)}{p_1^2-p_2^2} \kappa_{1,2}
-\frac{p_1^2-(p_1 p_2)}{p_1^2-p_2^2}  \kappa_{1,1}  \biggr)
\nonumber \\ 
&\hspace*{-55mm} + & \hspace{-30mm} 
\!\!(n\!-\!2) \!\biggl( 
\frac{p_1^2\!\!+\!\!m^2}{p_1^2 (p_1^2\!\!-\!\!p_2^2)} \kappa_{1,1}
\!\!-\!\frac{p_2^2\!\!+\!\!m^2}{p_2^2 (p_1^2\!\!-\!\!p_2^2)} \kappa_{1,2}
\!\!+\!\frac{m^2}{p_1^2 p_2^2} \widetilde{\kappa}\!\biggr)  \!\!\biggr] ,
\nonumber \\
\left.\lambda_3^{(1b)}\!(p_1^2,p_2^2,p_3^2)\!\right|_{\xi=0} 
\hspace*{-5mm} &=& \hspace*{-3mm}
\frac{g^2\eta C_A}{(4\pi)^{n/2}} \!\frac{(n\!-\!1) m}{2 (p_1^2\!-\!p_2^2)}
 (\kappa_{1,1}\!\!-\!\kappa_{1,2}),
\nonumber \\
\left.\lambda_4^{(1b)}\!(p_1^2,p_2^2,p_3^2)\!\right|_{\xi=0}
\hspace*{-5mm} &=& \hspace*{-3mm} 0 .
\end{eqnarray}
In fact, in an arbitrary gauge ($\xi\neq 0$), $\lambda_4^{(1b)}$ does
not vanish. It is proportional to $m\xi$.

\subsection{Transverse part}

In an arbitrary gauge, the $\tau$'s are not so compact as $\lambda$'s.
As an example, we show the result for one of them in the
Feynman gauge
\begin{eqnarray}
\left.\tau_1^{(1b)}\!(p_1^2,p_2^2,p_3^2)\!\right|_{\xi=0} 
\hspace*{-5mm} &=& \hspace*{-3mm}
- \frac{g^2\eta C_A}{(4\pi)^{n/2}}
\frac{(n - 1) m}{4 {\cal{K}}}  
\nonumber  \\ 
&\hspace*{-38mm}& \hspace{-33mm} \times \!
\biggl( 2 [(p_1 p_2)+m^2] \varphi_1
-\kappa_{1,2}-\kappa_{1,1} +2 \kappa_{0,3}
\nonumber \\ 
&&\hspace*{-3mm} -
(p_1\!-\!p_2)^2\; \frac{\kappa_{1,1}\!-\!\kappa_{1,2}}{p_1^2 \!-\! p_2^2}
\biggr) .
\end{eqnarray}
Results for all other functions are listed in \cite{DOS}.
 

In the limit of massless quarks, $m\to0$, the results simplify
considerably. 
First of all, in this limit, 
$\tau_1^{(1)}=\tau_4^{(1)}=\tau_5^{(1)}=\tau_7^{(1)}=0$
(this holds also in an arbitrary covariant gauge).
Furthermore, $\varphi_1\to\varphi$, $\varphi_2\to\varphi$,
$\kappa_{1,i}\to\kappa_{0,i}$, $\kappa_{2,i}\to\kappa_{0,i}$
and $\widetilde{\kappa}\to0$. 

\subsection{Renormalization}

In the limit $n\to 4$ ($\ep\to 0$) the only function in the quark-gluon
vertex which has an ultraviolet (UV) singularity {\em at one loop}
is the $\lambda_1^{(1)}$ function. 
Absence of UV-singularities in all other functions $\lambda_i^{(1)}$
($i>1$) and $\tau_i^{(1)}$ was one of the checks on our results.
In an arbitrary covariant gauge, the UV-singular part 
of $\lambda_1^{(1)}$ reads
\[
\frac{g^2\eta}{(4\pi)^{2-\ep}}
\left[ (1-\xi) C_F + \frac{1}{4} (4-\xi) C_A \right]
\left( \frac{1}{\ep} + \ldots \right) .
\]

To renormalize the results, 
we need to subtract the $1/\ep$'s, (possibly) getting some constant $R$
instead, 
$(1/\ep + \ldots ) \rightarrow \left( R + \ldots \right)$,
which depends on the renormalization scheme.
In the $\overline{\mbox{MS}}$ scheme $R=0$, because (see Eq.~(\ref{eta}))
$\eta = e^{-\gamma\ep} \left[ 1 + {\cal{O}}(\ep^2) \right]$ , 
so that $e^{-\gamma\ep}$ and $(4\pi)^{\ep}$ are absorbed by the
$\overline{\mbox{MS}}$ re-definition of the coupling constant $g^2$.

\section{SPECIAL LIMITS}

A few limits are of special interest: 
(i) the on-shell limit $p_1^2=p_2^2=m^2$ (with or without assuming
that the vertex function is being sandwiched between Dirac spinors);
(ii) the zero-momentum limit, when the gluon momentum vanishes ($p_3=0$);
(iii) the symmetric case, when $p_1^2=p_2^2=p_3^2$.
Since in all these cases we can put $p_1^2=p_2^2\equiv p^2$,
we start by considering this as the first step towards all these
limits.

In the case $p_1^2=p_2^2\equiv p^2$ some of the tensor structures
in the quark-gluon vertex become linearly dependent,
namely: $L_{2,\mu}$ and $T_{2,\mu}$, $L_{3,\mu}$ and $T_{1,\mu}$,
$T_{4,\mu}$ and $T_{7,\mu}$. 
Moreover, according to Eq.~(\ref{l4t6}), $\lambda_4$ and
$\tau_6$ vanish.
Therefore, the quark-gluon vertex
in this limit can be written as [cf. Eqs.~(\ref{lambda23t})
and (\ref{tau7t})]
\begin{eqnarray}
\label{Gamma12}
\left. \Gamma_{\mu}\right|_{p_1^2=p_2^2\equiv p^2} 
= L_{1,\mu} \lambda_1
+ L_{2,\mu} \widetilde{\lambda}_2
+ L_{3,\mu} \widetilde{\lambda}_3
\nn \\
+ T_{3,\mu} \tau_3  + T_{5,\mu} \tau_5  
+ T_{7,\mu} \widetilde{\tau}_7
+ T_{8,\mu} \tau_8 .
\end{eqnarray}
Only the $L_{1,\mu}$ contribution 
remains non-transverse in this limit. 

\subsection{On-shell limit}

The tensor structure of (\ref{Gamma12}) does not change
when we put $p^2=m^2$, without assuming that the vertex is 
sandwiched between Dirac spinors. If we keep $n$ as an arbitrary
parameter, the limit $p^2\to m^2$
is regular for all scalar functions involved in Eq.~(\ref{Gamma12}).
For example,
\begin{eqnarray}
\lambda_1^{(1)}(m^2,m^2,p_3^2) =
{\textstyle{1\over4}}
\frac{g^2\eta}{(4\pi)^{n/2}}\;   
\bigg\{ (2-\xi) C_A \kappa_{0,3}
\nonumber \\
+ \frac{n-2}{n-3} 
\left[ 2 (1-\xi)C_F + C_A \right] {\widetilde{\kappa}}
\bigg\} . \;
\end{eqnarray}
An interesting feature is that in this limit the integral
$J_2(1,1,1)$ reduces to the two-point function (with a factor
of $(n-4)^{-1}$ in front), whereas $J_1(1,1,1)$
remains non-trivial.

If we recall that the ``physical'' quark-gluon vertex should be
sandwiched between physical states obeying the Dirac equation,
then we arrive at
\begin{eqnarray}
\overline{u}(-p_1) \Gamma_{\mu} u(p_2) = 
F_1(p_3^2) \; \overline{u}(-p_1) \gamma_{\mu} u(p_2)
\nonumber \\
- {\textstyle{1\over2m}} F_2(p_3^2) \;
\overline{u}(-p_1) \sigma_{\mu \nu}\, p_3^\nu u(p_2),
\end{eqnarray}
where $F_1(p_3^2)$ and $F_2(p_3^2)$ are often called the Dirac and Pauli
form factors, respectively.
In terms of the (modified) longitudinal and transverse functions we get
\begin{eqnarray}
F_1+F_2 \!\! &=& \!\!\lambda_1 + p_3^2 \tau_3 -2m \tau_5 +
\half ( p_3^2 \!-\! 4 m^2) \tau_8 ,
\nn \\
{\textstyle{1\over2m}} F_2 \!\! &=& \!\! -2 m \widetilde{\lambda}_2
+ \widetilde{\lambda}_3 \!-\! \tau_5 \!+\! \half p_3^2 \widetilde{\tau}_7
- m \tau_8 . 
\end{eqnarray}  

\subsection{Zero-momentum limit}

Let us now consider the off-shell
zero-momentum limit $p_3=0$ ($p_2=-p_1\equiv p$).
Upon putting $p_1^2=p_2^2=p^2$, the next step is
to put $p_3^2=0$ (which implies $p_3=0$). 
In this limit,
most of the transverse structures vanish and the quark-gluon vertex
looks like
\begin{eqnarray}
\label{qqg_zerom}
\left. \Gamma_{\mu}\right|_{p_2=-p_1\equiv p,\; p_3=0}
\hspace*{-3mm}&=& \hspace*{-3mm}L_{1,\mu} \lambda_1
+ L_{2,\mu} \widetilde{\lambda}_2
+ L_{3,\mu} \widetilde{\lambda}_3
\nn \\
&\hspace*{-3mm}=& \hspace{-3mm}
\gamma_{\mu} \lambda_1
+ 4p_{\mu}\pslash \widetilde{\lambda}_2
- 2p_{\mu}  \widetilde{\lambda}_3.
\end{eqnarray}
Of course, all integrals can be reduced to two-point
functions and tadpoles. 
The corresponding scalar functions are regular in this limit.

In the massless case ($m=0$), only two relevant structures 
remain in (\ref{qqg_zerom}). The corresponding results have
been compared with \cite{BL}.

\subsection{Symmetric case}

The limit $p_1^2=p_2^2=p_3^2\equiv p^2$ is interesting for studying
the $Z$-factors and renormalization group quantities
in the MOM scheme. Usually, a Euclidean symmetric point 
is considered, $p^2=-\mu^2$.
For massless quarks, the one-loop calculation of the
quark-gluon vertex was performed in \cite{PT} (see also
in \cite{CG}). For massive quarks, one of the scalar functions,
the coefficient of $\gamma_{\mu}$, was presented in \cite{DTP}.
We confirm all these results, and also have obtained 
expressions for the remaining functions (for details, see
in \cite{DOS}).

\section{CONCLUSIONS}

We have reviewed results for the one-loop quark-gluon
vertex in an arbitrary covariant gauge and space-time dimension. 
The calculation was carried out with massive quarks.

We decomposed the quark-gluon vertex into
longitudinal and transverse parts (\ref{LT}) 
(like the decomposition in QED \cite{BC1}). 
In general, 12 scalar functions (four $\lambda$'s and eight
$\tau$'s) are needed to define the quark-gluon vertex. 
In particular, we have found that
the function $\lambda_4$, the coefficient of 
$\sigma_{\mu \nu}(p_1-p_2)^{\nu}$, which does not appear in QED
(at least, at the one-loop level),
does not vanish in QCD. Moreover, it contributes to the non-Abelian 
sector of the WST identity (\ref{WST_qqg}). 
Various special cases of the general results were compared with those
of Refs.~\cite{BC1,KRP,BKP,PT,CG,DTP,ALV} (see in \cite{DOS}).

Starting from the general off-shell expressions 
in an arbitrary space-time dimension, $n$, for the longitudinal and transverse
parts of the vertex, we have derived results for the on-shell
limits which also are valid for an arbitrary value of the space-time
dimension.
For special cases, our on-shell results have been compared against
those from Refs.~\cite{BL,NPS,Muta,Bailin+Love}.

We have calculated other functions involved and
checked that our results obey 
the WST identity (\ref{WST_qqg}),
for arbitrary $n$ and $\xi$.

In principle, some techniques which can be used for the calculation
of the two-loop off-shell quark-gluon vertex, at least in the $m=0$ case,
are already available \cite{UD,Tar}, although the problem of
higher powers of irreducible numerators is still difficult
for algorithmization. For special limits, the calculation is
very similar to the three-gluon vertex, which was calculated
at two loops in \cite{DOT2} (the zero-momentum limit) and in 
\cite{DO1} (the on-shell case).

\vspace{3mm}

{\bf Acknowledgements.} 
It is a pleasure to thank the organizers of `Loops and Legs 2000'
for creating a stimulating atmosphere at the conference.
We are grateful to O.V.~Tarasov for useful discussions.
This research has been supported by the Research Council of Norway.
A.~D.'s research was supported by the Alexander von Humboldt
Foundation and by DFG.
Research by L.~S. was supported by the Norwegian State Educational 
Loan Fund.

\end{document}